\documentclass[twocolumn,preprintnumbers,amsmath,amssymb]{revtex4}
\usepackage{latexsym,amssymb,amsthm,amsmath,epsfig}

\begin{document}

\title{Creation of $p$-wave Feshbach molecules in the selected angular momentum states using an optical lattice}
\author{Muhammad Waseem}
\email{waseem@ils.uec.ac.jp}
\author{Zhiqi Zhang}
\author{Jun Yoshida}
\author{Keita Hattori}
\author{Taketo Saito}
\author{Takashi Mukaiyama}
\affiliation{Department of Engineering Science, University of Electro-Communications, Tokyo 182-8585, Japan}
\affiliation{Institute for Laser Science, University of Electro-Communications, 1-5-1Chofugaoka, Chofu, Tokyo 182-8585, Japan}
\date{\today }

\begin{abstract}
We selectively create $p$-wave Feshbach molecules in the $m_{l}=\pm 1$ orbital angular momentum projection state of $^{6}$Li. We use an optical lattice potential to restrict the relative momentum of the atoms such that only the $m_{l}=\pm 1$ molecular state couples to the atoms at the Feshbach resonance. We observe the hollow-centered dissociation profile, which is a clear indication of the selective creation of $p$-wave molecules in the $m_{l}=\pm1$ states. We also measure the dissociation energy of the $p$-wave molecules created in the optical lattice and develop a theoretical formulation to explain the dissociation energy as a function of the magnetic field ramp rate for dissociation. The capability of selecting one of the two closely-residing $p$-wave Feshbach resonances is useful for the precise characterization of the $p$-wave Feshbach resonances.
\end{abstract}

\maketitle

\section{introduction}

Fermionic $p$-wave superfluidity presents a rich variety of novel phenomena caused by complex order parameters. Recent experimental advances in controlling interactions in ultracold atomic gases using Feshbach resonances have awakened expectations for realizing $p$-wave superfluidity of fermionic atoms, which would offer great opportunities to study superfluid phases with the precise control of experimental parameters. While fermions with $s$-wave interactions show smooth crossover when the interaction strength is varied across a Feshbach resonance~\cite{reg, bartenstein, zw}, theoretical calculations predict quantum phase transitions across a $p$-wave Feshbach resonance~\cite{vg}. The phase transition from $p_x$ superfluid to a time-reversal breaking $p_x + i p_y$ superfluid has been discussed \cite{vg, Cheng}. The system has also been predicted to show topological quantum phase transition from a gapped to a gapless $p_x + i p_y$ superfluid state. An anisotropic feature of $p$-wave interaction plays important roles in the emergence of such various superfluid phases.

Due to dipole-dipole interactions between two atoms in a $p$-wave Feshbach molecular state, $p$-wave Feshbach resonances possess doublet structures originating from the different angular momentum projections $m_{l}=0 $ and $m_{l}=\pm 1$~\cite{ticknor}. The splitting of Feshbach resonance is known to have significant influence on the resultant superfluid phases~\cite{vg}. While the superfluid phase diagram becomes simple for the case of a large resonance splitting, a rich and complicated superfluid phase diagram is expected to appear for the case of a small resonance splitting due to varying anisotropy of the interactions which depends sensitively on the magnetic field~\cite{vg}. 

Since an atomic loss near $p$-wave Feshbach resonances is a main obstacle to achieve $p$-wave superfluid, elastic and inelastic collisional properties of the atoms near $p$-wave Feshbach resonances have been investigated intensively using $^{40}$K and $^{6}$Li atoms~\cite{zhang, shunk, gab, inada, fuch, naka, raw}. Since $^{40}$K has a relatively large resonance splitting on the order of 0.5~G~\cite{reg3, gun, gab, thy}, the JILA group selectively created $p$-wave molecules in both the $m_{l}=\pm 1$ states and the $m_{l}=0$ state~\cite{gab}. 
In contrast, the splitting of the two resonances in $^{6}$Li is predicted to be quite small~\cite{chevy}, and therefore, the splitting of the two resonances has not been resolved in $^{6}$Li yet. 
Since the small splitting is an essential origin of the rich superfluid phase diagram, $^{6}$Li is an attractive system to study a $p$-wave superfluid. However, the small splitting makes the experimental characterization of $p$-wave Feshbach resonances quite challenging because of the incapability of accessing each resonance independently. Actually in the previous experiments of the creation of the $p$-wave Feshbach molecules in $^{6}$Li, the selection of the angular momentum projection of the $p$-wave molecules was not possible because of the small resonance splitting.

This paper presents the selective creation of $p$-wave Feshbach molecules of $^{6}$Li atoms in the $m_{l}=\pm 1$ states by using an optical lattice potential. We used a one-dimensional optical lattice whose laser beam propagates along the quantization axis to create a two-dimensional gas of atoms. As a result of the tight confinement of the atoms along the lattice direction, the coupling of the atoms to the $p$-wave molecular state of $m_{l}=0$ is suppressed and the coupling to the $m_{l}=\pm 1$ states is singled out despite the small splitting between $m_{l}=0$ and $m_{l}=\pm 1$ resonances. In this configuration, $p$-wave molecules in $m_{l}=\pm 1$ are created by adiabatically ramping the magnetic field across Feshbach resonance. To reveal the symmetry of the $p$-wave molecules, the momentum distribution of the atoms dissociated from the molecules was captured, showing a hollow-centered profile. We also studied the dissociation dynamics of the $p$-wave molecules purely in the $m_{l}=\pm 1$ states and the mixture of the $m_{l}=\pm 1$ and $m_{l}=0$ states. We developed a theoretical formulation to explain the dissociation energy as a function of the magnetic field ramp rate for molecular dissociation. The measured sub-linear dependence of the dissociation energy on the field ramp rate can be quantitatively explained by the solution of the rate equation.

The rest of the paper is organized as follows. In Sec. IIA, we introduce the experimental setup for the preparation of a two-dimensional Fermi gas of $^{6}$Li atoms. Section IIB contains the study on the creation of $p$-wave molecules in a two-dimensional trap and the observation of the dissociation profile, which proves the selective $p$-wave molecular creation in the $m_{l}=\pm 1$ states. Section IIC describes the study on the molecular dissociation energy. Finally the conclusion and future outlook are summarized in Sec. III.

\section{experiment}
\subsection{Preparation of a two-dimensional gas of $^{6}$Li atoms}

We prepared a degenerate Fermi gas of $^{6}$Li atoms in the hyperfine ground state of $\left\vert F,m_{F}\right\rangle= \left\vert 1/2,1/2\right\rangle (\equiv \left\vert 1\right\rangle )$ and $\left\vert F,m_{F}\right\rangle =  \left\vert 1/2,-1/2 \right\rangle (\equiv \left\vert 2\right\rangle )$ by employing all optical methods described in detail elsewhere \cite{Inada_Tc}. Briefly, we performed evaporative cooling at 300~G in a single-beam optical dipole trap with the beam waist of 34~$\mu$m propagating along the $y$ direction. At this magnetic field, the absolute value of the scattering length between the states $\left\vert 1\right\rangle$ and $\left\vert 2\right\rangle$ lies at the local maximum. We obtained $2 \times 10^{6}$ atoms at a temperature of $T/T_{\rm F}= 0.13$. At the final trap depth of the evaporation, trap frequencies were $(\omega_{x}, \omega_{y}, \omega_{z}) = 2 \pi \times (478, 10, 660)$~Hz. After the preparation of the two-component Fermi degenerate gas, we shined the resonant light with a duration of 50~$\mu$s to remove the atoms in the $\left\vert 2\right\rangle$ state. We confirmed that all atoms in $\left\vert 2\right\rangle$ were removed by applying the conventional Stern-Gerlach measurement.

\begin{figure}[tbp]
\includegraphics[width=3.4 in]{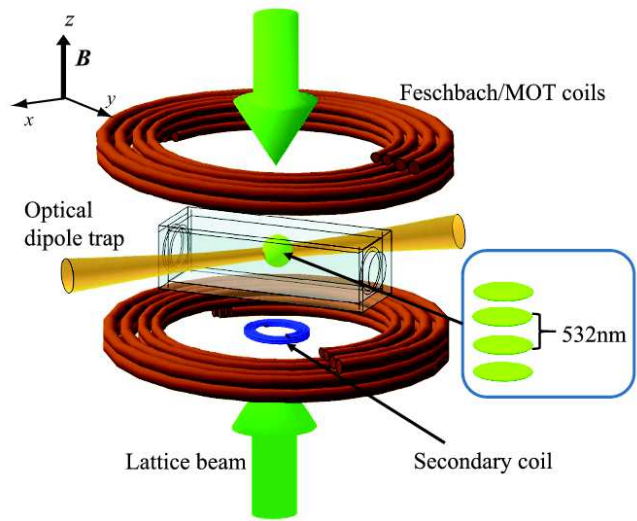}
\caption{Experimental setup. A linearly polarized laser for an optical lattice was incident along the $z$ axis onto the atoms trapped by a single-beam optical dipole trap. The lattice laser was retro-reflected to create a one-dimensional optical lattice potential, and the homogeneous Feshbach magnetic field was also applied along the $z$ axis. The small secondary coil was placed beneath the glass cell, which was used for the fast magnetic field ramp for the molecular dissociation in the latter experiment. }
\label{fig1}
\end{figure}

To prepare a two-dimensional Fermi gas, we shined another laser at 1064~nm propagating along the $z$ direction and retro-reflect the laser to create a one-dimensional optical lattice potential as shown schematically in Fig.~\ref{fig1}. To realize a perfect overlap between the incoming and retro-reflecting beams, we pulsed the optical lattice lasers for a short time scale (pulse duration of 1~$\mu$s) and observed the Kapitza-Dirac scatterings~\cite{KDS}. We maximized the Kapitza-Dirac scattering efficiency at a fixed pulse duration by optimizing the position of the retro-reflection mirror. From the measurement of the trap frequencies in the $x$ and $y$ directions with and without the retro-reflection beam, we determined the beam waist of the optical lattice laser to be 73~$\mu$m. Although, in an ideal condition, the optical lattice depth $V_{\rm lat}$ is four times the trap depth of the dipole force applied by the incoming beam $V_{\rm dip}$, the actual optical lattice depth is described as $V_{\rm lat}=\alpha \times V_{\rm dip}$, where $\alpha < 4$ because of the imperfection of the retro-reflection due to the loss of laser intensity or polarization rotation in the reflected beam. From the information of the trap frequencies with and without the retro-reflection beam, we determined $\alpha=2.2 \pm 0.2$ in our experimental setup. The optical lattice depth was calibrated to be $V_{\rm lat}= 3.8$~$E_{\rm rec}$ at 350~mW of the incoming laser intensity, where $E_{\rm{rec}}=\hbar^2 k^2/(2m)\approx k_{\rm B} \times 1.4$~$\mu$K is the recoil energy of $^{6}$Li.

\begin{figure}[tbp]
\includegraphics[width=3.4 in]{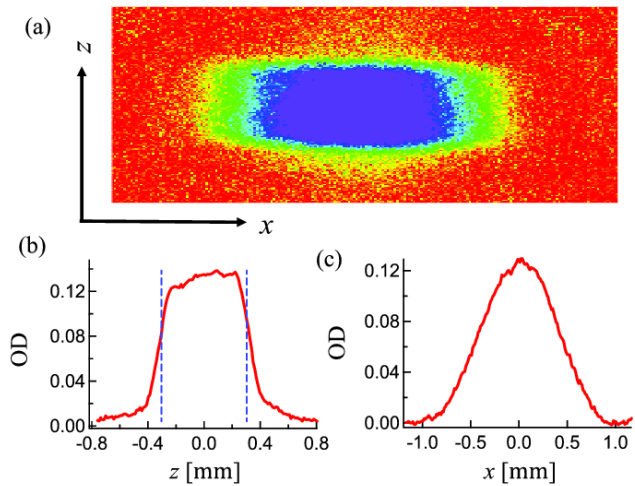}
\caption{Brillouin zone mapping for the optical lattice depth of $9.3 E_{\rm rec}$. (a) Absorption image was taken after 5~ms expansion after adiabatically ramping down the optical lattice. (b) Integrated one-dimensional profile along the $z$ direction (lattice direction). The  distribution shows a square shape, indicating that the atomic population was restricted to the first Brillouin zone. The vertical dashed lines show the position where the band edge was expected to be located. (c) Integrated one-dimensional profile along the $z$ direction, which shows a nearly Gaussian profile.}
\label{fig2}
\end{figure}

At the lattice laser power of 350~mW, $T/T_{\rm F}$ was determined to be 0.4. The temperature of the gas normalized by the Fermi temperature $T/T_{\rm F}$ was determined from the momentum distribution in the $x$-$y$ plane using a two-dimensional Thomas Fermi profile~\cite{fro}. The Fermi temperature was $T_{\rm F}=\hbar \omega_{r} \sqrt{2N}/k_{\rm B} = 1.7 \mu{\rm K}$, where $\omega_{r}=\sqrt{\omega_{x} \omega_{y}}$ is the radial trap frequency and $N$ is thenumber of atoms per site (here, $5\times 10^3$). Since the axial excitation energy $\hbar \omega_{z}\approx k_{\rm B} \times 5.5$~$\mu$K with $\omega_{z} = 2 \pi \times 115$~kHz is higher than both $ k_{\rm B} T_{\rm F} $ and $k_{\rm B} T$, most of the atoms are in the motional ground state along the $z$ axis. The fraction of the number of atoms in the first motional excited state to the number of atoms in the motional ground state is determined by ${\rm exp}(-\hbar \omega_z/2 k_{\rm B} T) \approx 0.02$. Therefore 2~\% of the atoms would reside in the first motional excited state.
To verify that the major fraction of the atoms are in the motional ground state along the $z$ direction, we performed the Brillouin zone mapping~\cite{Greiner, kohl}. After the preparation of the atoms in the two-dimensional trap, we ramped down the optical lattice depth within 0.5~ms, which was slow enough to guarantee that the atoms stayed in the lowest band with the quasi-momentum unchanged. After the ramp down, we released the cloud and let it expand freely for 5~ms before taking the absorption image along the $y$ axis. 
Figure~\ref{fig2}(a) shows a typical absorption image taken after ramping down the optical lattice for the case of 9.3~$E_{\rm{rec}}$ lattice depth. Figures~\ref{fig2}(b) and \ref{fig2}(c) show the integrated one-dimensional profile along the $z$ and $x$ axes, respectively. The profile along the $z$ axis shows a square distribution, which indicates that all atoms were in the first Brillouin zone and no higher Brillouin zones were occupied by the atoms. In contrast, the profile along the $x$ axis shows a Gaussian profile (Fig.~\ref{fig2}(c)).

\subsection{$p$-wave Feshbach molecules in $m_{l}=\pm 1$ state}

The $p$-wave Feshbach resonances are known to possess doublet structures arising from the energy splitting of closed-channel molecular levels caused by magnetic dipole-dipole interaction~\cite{ticknor}. The resonance magnetic field depends on $m_{l}$, the projection of the angular momentum onto the quantization axis defined by the direction  of the external magnetic field. In the case of $^{6}$Li atoms, the splitting between two resonances for $m_{l}=0$ and $m_{l}=\pm 1$ has not been resolved experimentally so far and the splitting is considered to be quite small~\cite{chevy}. This is clearly in contrast with the case of $^{40}$K, where the resonance splitting is larger than the resonance width and the $p$-wave Feshbach resonances for $m_{l}=0$ and $m_{l}=\pm 1$ can be addressed separately by tuning the magnetic field. Here, we used a one-dimensional optical lattice potential to tightly confine the $^{6}$Li atoms along the quantization axis (referred to as a 2D trap) to single out the $p$-wave Feshbach resonance for the $m_{l}=\pm 1$ state. Since all atoms are in the motional ground state along the $z$ direction, atoms can only have a relative momentum along the $x$ or $y$ direction. Then, atoms can only collide through $p$-wave collisions with $m_{l}=\pm 1$ symmetry.

\begin{figure}[tbp]
\includegraphics[width=3.2 in]{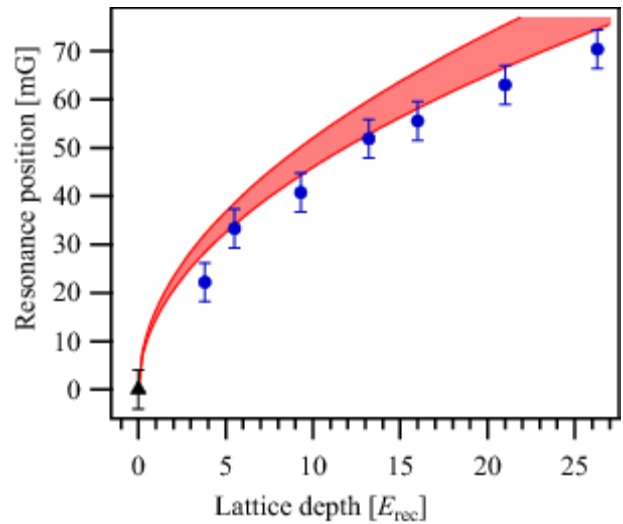}
\caption{Solid circles show the measured shift in Feshbach resonance versus optical lattice depths.  The resonance location was determined from the sharp lower field edge of the atomic loss profile. The error bars indicate an uncertainty in the determination limited by the stability of the magnetic field. Solid triangle shows the resonance position with no optical lattice confinement. Solid curve shows the calculated resonance shift from the collision energy of the atoms.}
\label{fig3}
\end{figure}

Once the 2D Fermi gas was prepared at 300~G, we swept the magnetic field to 159~G, where the $p$-wave Feshbach resonance for the atoms in the $\left\vert 1 \right\rangle$ state was located. We stabilized the current running in the Feshbach coils and reduced the magnetic field fluctuation to less than 8~mG~\cite{naka}. We determined the $p$-wave Feshbach resonance position from the sharp lower field edge of the atomic loss profile~\cite{naka}. 
In the 2D trap, we observed a shift in Feshbach resonance toward the higher magnetic field side to the original resonance due to the optical lattice confinement. The experimentally measured resonance shift versus optical lattice depth is plotted with blue markers in Fig.~\ref{fig3}. The red shaded curve shows the calculated shift based on the model in which the collision energy of the atoms was determined by the kinetic energy (namely the Fermi energy in this case) of the atoms along the $x$ and $y$ directions \cite{gun}. The width of the shade indicates the uncertainty of the magnetic moment of the molecular state. We found reasonable agreement between the model and the experimental result, but some systematic deviation is still visible.
One possible origin of the deviation is the interatomic interactions enhanced near the $p$-wave Feshbach resonance. Resonance shift due to combined effect of tight confinement and the $s$-wave interatomic interactions has been reported by Syassen et al. \cite{Syassen} in the system of bosons in a three dimensional optical lattice. There may be a way to generalize the idea to the system of fermions with $p$-wave interactions. More detailed understanding of this shift is for our future study.

We created $p$-wave molecules in the 2D trap by adiabatically ramping the magnetic field from above to below the Feshbach resonance for the $\left\vert 1 \right\rangle$ state similar to our previous work~\cite{inada}. The final value of the magnetic field sweep was chosen to optimize the molecular creation efficiency. Since the resonance magnetic field depends on the optical lattice depth, the final value of the magnetic field sweep was chosen differently for different lattice depth conditions. The optimum magnetic field value roughly lies 25~mG below the resonance, which is also consistent with the previous observations~\cite{inada}. The molecule conversion efficiency was approximately 15~\%, which decreases with increasing lattice depth presumably because of faster loss of molecules due to atom-molecule inelastic collisions for higher atomic density condition. After the molecular creation, we immediately removed unpaired atoms by shining the resonance light pulse for 100~$\mu$s. After removing the unpaired atoms, we ramped up the magnetic field above the resonance to convert molecules back into atoms, and captured absorption images. We carefully checked the quality of the blast light by intentionally keeping the magnetic field away from the resonance and confirmed that all atoms were removed.

\begin{figure}[tbp]
\includegraphics[width=3.4 in]{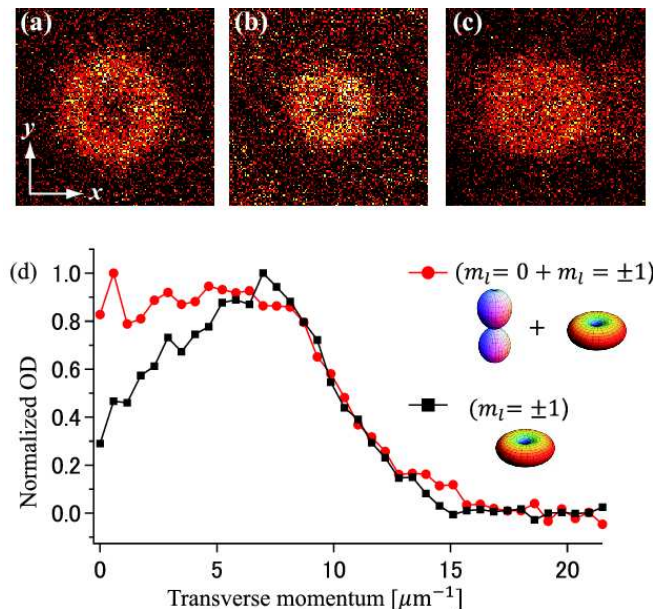}
\caption{Radial momentum distribution of the atoms after the molecular dissociation, taken at the lattice depth of (a) $V_{\rm lat}=3.8 E_{\rm rec}$ and (b) $V_{\rm lat}=5.5 E_{\rm rec}$, respectively. The image (c) was taken without the optical lattice potential. Images (a) and (c) were taken with 1.3~ms time-of-flight with the magnetic field ramp rate of 9~G/ms, and image (b) was taken with 1.1~ms time-of-flight with the magnetic field ramp rate of 5~G/ms. Each image is an average of 5 to 10 shots. The hollow-centered dissociation profile shown in (a) was due to the selective creation of the $p$-wave molecules in the $m_{l}=\pm 1$ states. (d) Radially averaged profile of the images for the atoms with and without the lattice potential are shown with black squares ($V_{\rm lat}=3.8 E_{\rm rec}$) and red circles, respectively. Clear dip at the low dissociation momentum region is visible in the plot of black squares. As shown graphically in the inset, molecules created in the lattice are purely in the $m_{l}=\pm 1$ states and those created without lattice potential are the mixture of $m_{l}=\pm 1$ and $m_{l}=0$ $p$-wave molecules.}
\label{fig4}
\end{figure}

We confirmed that we selectively create $p$-wave molecules in the $m_{l}=\pm 1$ states by measuring a momentum distribution of the atoms dissociated from the molecules. To observe the dissociation profile, we turned off the optical lattice and released the molecules to let them ballistically expand, and then, immediately ramped up the magnetic field across the Feshbach resonance within 50~$\mu$s to make the molecules dissociate. After 1.1 to 1.3~ms of ballistic expansion, we took an absorption image along the quantization axis. Figures~\ref{fig4}(a) and \ref{fig4}(b) show the dissociation momentum profiles of the atoms from the $p$-wave Feshbach molecules at an optical lattice depth of (a) $V_{\rm lat} = 3.8 E_{\rm rec}$ (with 1.3~ms expansion and 9~G/ms dissociation field ramp) and (b) $V_{\rm lat} = 5.5 E_{\rm rec}$ (with 1.1~ms expansion and 5~G/ms dissociation field ramp). The dissociation profiles clearly show holes at the center in the time-of-flight images. This is a clear signature of the selective molecular creation in the $m_{l}=\pm 1$ states, which has also been observed in the $^{40}$K system by taking advantage of the large Feshbach resonance splitting between $m_{l}=\pm 1$ and $m_{l}=0$ states~\cite{gab}. In $^6$Li the Feshbach resonance splitting is too small to be resolved, but we were still able to selectively create $p$-wave molecules in the $m_{l}=\pm 1$ states with the help of tight confinement due to an optical lattice potential. 

However, we needed an experimental confirmation that the hollow-centered dissociation profile is genuinely due to the selective molecular creation in the $m_{l}=\pm 1$ states. Even if the molecules are created equally in the $m_{l}=0$ and $m_{l}=\pm 1$ states, such a hollow-centered dissociation profile can also arise from dissociation kinetics~\cite{durr}. When we ramp up the magnetic field fast enough such that the added dissociation energy is much larger than the kinetic energy of the molecules, the dissociation profile can be modified drastically from a Gaussian profile and can sometimes show a mono-energetic spherical expansion, which ends up with a hollow-centered dissociation profile in an absorption image. To make sure that the hollow-centered dissociation profile we observed is genuinely due to the selective creation of $p$-wave molecules in the $m_{l}=\pm 1$ states, we performed the same experiment without an optical lattice potential. For this confirmation, we changed the polarization of the retro-reflected beam of the optical lattice laser to realize a cross-beam trap configuration (referred to as a 3D trap). Figure~\ref{fig4}(c) shows the dissociation profile of the molecules without an optical lattice potential. The image was taken with an expansion time of 1.3~ms and the magnetic field ramp rate of 9~G/ms, which are the same as the conditions for the data shown in Fig.~\ref{fig4}(a). In Fig.~\ref{fig4}(c), we observe a relatively uniform dissociation profile, which shows a clear contrast to the result shown in Fig.~\ref{fig4}(a). Figure~\ref{fig4}(d) shows the results of the radial averaging of the dissociation profiles shown in Figs.~\ref{fig4}(a) (black squares) and \ref{fig4} (c) (red circles). Again, we can clearly see that the data taken with the lattice potential shows a dip at the center of the dissociation momentum profile. In contrast, we see a rather flat-top profile in the data without the lattice potential in a low-momentum region. From the difference in the dissociation profiles with and without an optical lattice potential, we can conclude that the hollow-centered dissociation profile observed in this measurement is due to the selective creation of $p$-wave molecules in the $m_{l}=\pm 1$ states.

\subsection{Measurement of molecular dissociation energy}

\begin{figure}[tbp]
\includegraphics[width=3.2 in]{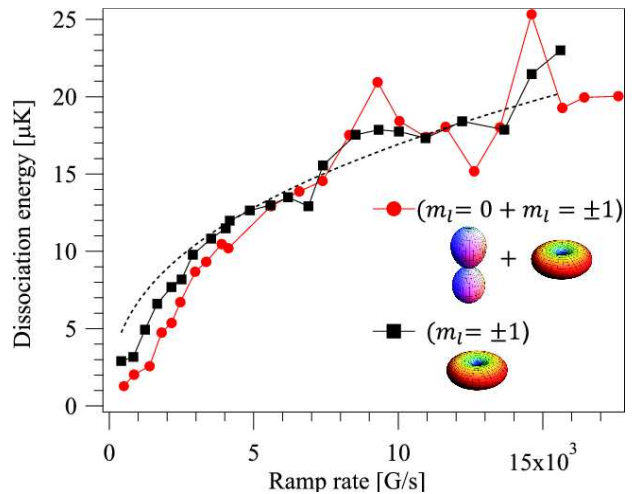}
\caption{Dissociation energy as a function of the magnetic field ramp rate. Red circles represent the dissociation energy in the 3D trap and the black squares represent the dissociation energy at lattice depth of $V_{\rm lat}=8.4E_{\rm rec}$. Dashed curve is the result of fitting with Eq.~(\ref{en}).}
\label{fig5}
\end{figure}

We further investigated the dissociation dynamics of $p$-wave Feshbach molecules. More specifically, we measured the dissociation energy of the $p$-wave molecules created in 2D and 3D traps. The dissociation energy can be determined by taking the sum of the atomic energies at all pixels in the absorption images. The measured dissociation energies as a function of the magnetic field ramp for the molecular dissociation are shown in Fig.~\ref{fig5}. The data for the atoms in the 2D trap at the lattice depth of $V_{\rm lat}=8.4E_{\rm rec}$ are shown with the black squares, and those for the atoms in the 3D trap are shown with red circles. When the $p$-wave molecules were created in the optical lattice potentials, the created molecules were only in the $m_l =\pm 1$ states. In the 3D trap on the other hand, the $p$-wave molecules were created equally in both the $m_l = 0$ and $m_l = \pm 1$ states. In the process of molecular dissociation, a $p$-wave molecule in $m_l = 0$ dissociates into two atoms flying in $z$ direction and a $p$-wave molecule in $m_l = \pm 1$ dissociate into two atoms flying in $x$-$y$ plane. Therefore, in the 3D trap, molecules dissociate into atoms flying isotropically in all three directions. Since summing up the energy of the atoms in the absorption image taken along the $z$ direction only counts the energy in the $x$-$y$ plane, we underestimates the dissociation energy by a factor of 3/2 for the 3D trap case. In the analysis, we multiplied the factor of 3/2 to the obtained dissociation energy from the image to obtain the real dissociation energy for the molecules created in the 3D trap. As clearly seen, the dissociation energy of the molecules in the $m_l =\pm 1$ states and that of the mixture of molecules in the $m_l =0$ and $m_l =\pm 1$ states match with each other within the error of our measurement. The fact that the measured dissociation energies are the same for the molecules created in the 2D and 3D traps even with the different factors (unity for the 2D trap and 3/2 for the 3D trap) indicates that the $p$-wave molecules are created almost purely in the $m_l = \pm 1$ states in the 2D trap. This is consisnt with the fact that the number of atoms in the first motional excited state along the $z$ direction is only 2~\% of total number of atoms and the imperfection arising from this contribution is negligibly small.
The dissociation energy plots show the sub-linear dependence on the magnetic field ramp rate, indicating that the molecular decay is faster for the higher quasi-bound molecular energy. As long as the atoms are in the motional ground state along the $z$ direction, the dissociation energy does not depend on the lattice depth. The measurement of the dissociation is performed for the lattice depths of $V_{\rm lat}=3.8E_{\rm rec}$ and $V_{\rm lat}=5.5E_{\rm rec}$, and data for the both lattice depths overlap nicely with the plot for $V_{\rm lat}=8.4E_{\rm rec}$ shown in Fig.~\ref{fig5}.

The dissociation dynamics of the $p$-wave molecules can be quantitatively explained by the rate equations \cite{sensei}. We first consider that the initial population is entirely in the molecular state. Fraction of the molecules $f(\epsilon)$ at energy $\epsilon$ can be described as
\begin{equation}
\frac{df(\epsilon)}{d\epsilon}= \frac{df(\epsilon)}{dt} \frac{dt}{d\epsilon} = -\Gamma(\epsilon)  f(\epsilon) \frac{d t}{d \epsilon},
\label{rate}
\end{equation}
where $\Gamma(\epsilon)$ is the dissociation rate of the molecules at energy $\epsilon$. It has been discussed theoretically that the $p$-wave molecules above resonance decay into scattering atomic states with the rate of $\Gamma(\epsilon)= A \epsilon^{3/2}$ with $A = 2 \sqrt{m}/(k_{\rm e}  \hbar^2)$, where $k_{\rm e}$ is an effective range~\cite{victor}. This energy dependence of the decay rate has been experimentally confirmed in the measurement of the lifetime of the $p$-wave molecules above resonance by the JILA group~\cite{gab}.

The energy of the bare molecular state depends linearly on the magnetic field $B$ such that $\epsilon=\Delta\mu (B-B_{0})$, where $\Delta\mu$ is the relative magnetic moment of the molecular state to the atomic state and $B_{0}$ is the Feshbach resonance magnetic field. Then, Eq.~(\ref{rate}) can be rewritten as 
\begin{equation}
\frac{d f(\epsilon)}{d \epsilon}= -A \epsilon^{3/2} f(\epsilon) \left( \Delta\mu \left| \dot{B} \right| \right)^{-1}. 
\end{equation}
Solution of the above differential equation is 
\begin{equation}
f(\epsilon)= e^{-\alpha \epsilon^{5/2} }
\end{equation}
with $\alpha=\frac{2 A }{5 \Delta\mu \left| \dot{B} \right|}$. In laboratory frame, the average energy of atoms after dissociation becomes
\begin{equation}
\delta E=\int_{0}^{\infty} \frac{\epsilon}{2} [-df(\epsilon)]\\
=\frac{\Gamma(7/5)}{2} \left(\frac{5 k_{\rm e} \hbar^2 \Delta \mu |\dot{B}|}{4 \sqrt{m}}\right)^{2/5}
\label{en}
\end{equation}

Using $\Delta\mu=k_{\rm B} \times 113 \mu$K/G~\cite{fuch}, we fitted the experimental results with Eq.~(\ref{en}) by taking the effective range $ k_{\rm e}$ as a fitting parameter. Result of the fitting to the data with the black squares is shown by the black dashed curve in Fig.~\ref{fig5}. It is clearly seen that the experimental data shows $|\dot{B}|^{2/5}$ dependence as expected from Eq.~(\ref{en}). However, the obtained effective range is $k_{\rm e} = 0.14(1) a_0^{-1}$, which is approximately a factor of two larger than the value determined from the cross-dimensional thermalization measurement~\cite{naka}. It is unclear whether the expression of the dissociation rate $\Gamma (\epsilon) = 2 \sqrt{m}/(k_{\rm e}  \hbar^2) \epsilon^{3/2}$ is exact over the whole measurement condition, since this expression is valid only for the large scattering volume limit~\cite{victor}. Deeper quantitative understanding of the molecular dissociation is our next challenge.

\section{conclusion and outlook}
We successfully singled out one of the two closely-residing $p$-wave Feshbach resonances using a strong confinement. We demonstrated the selective creation of $p$-wave molecules in the $m_{l}=\pm 1$ state in $^{6}$Li by employing a one-dimensional optical lattice potential along the quantization axis. We confirmed that the created molecules are in the $m_{l}=\pm 1$ state from the measurement of the hollow-centered momentum distribution of the atoms dissociated from the molecules. We also measured the dissociation energy of the $p$-wave molecules as a function of the magnetic field ramp rate for molecular dissociation. The measured sub-linear dependence of the dissociation energy to the field ramp rate can be quantitatively explained by the solution of the rate equation. The way to single out one of the two closely-residing resonances using an optical lattice potential will be useful for the characterization of $p$-wave Feshbach resonance, such as the determination of the scattering parameters for the $p$-wave Feshbach resonances performed previously excluding the resonance splitting.

\section*{ACKNOWLEDGMENT}
We would like to thank Dr. Shinsuke Haze for his support in the experiment. This work is supported by a Grant-in-Aid for Scientific Research on Innovative Areas (Grant No. 24105006). MW acknowledges the support of a Japanese government scholarship (MEXT).

\end{document}